\documentclass{aastex}
\usepackage{epsfig}
\usepackage{emulateapj5}
\usepackage{natbib}

\newcommand{\bpic}{$\beta$~Pictoris}
\newcommand{\bp}{$\beta$~Pic}
\newcommand{\hh}{H$_{2}$}
\newcommand{\Os}{\ion{O}{6}}
\newcommand{\Ct}{\ion{C}{2}}
\newcommand{\cohh}{\mbox{CO$/$\hh}}
\newcommand{\kms}{km~s$^{-1}$}
\newcommand{\cm}{cm$^{-2}$}

\newcommand{\fuse}{{\it FUSE}}
\newcommand{\hst}{{\it HST}}
\newcommand{\aumic}{AU~Microscopii}
\newcommand{\au}{AU~Mic}

\shorttitle{Rapid Dissipation of H$_2$ from the AU~Mic Disk}
\shortauthors{Roberge et al.}

\begin{document}


\title{Rapid Dissipation of Primordial Gas from the AU~Microscopii Debris Disk}


\author{Aki Roberge\altaffilmark{1}, 
Alycia J.\ Weinberger\altaffilmark{1},
Seth Redfield\altaffilmark{2},
and Paul D.\ Feldman\altaffilmark{3}}

\email{akir@dtm.ciw.edu, alycia@dtm.ciw.edu, 
sredfield@astro.as.utexas.edu, pdf@pha.jhu.edu }


\altaffiltext{1}{Dept.\ of Terrestrial Magnetism, Carnegie Institution
of Washington, 5241 Broad Branch Road, NW, Washington, DC, 20015}
\altaffiltext{2}{Harlan J.\ Smith Postdoctoral Fellow, 
McDonald Observatory, University of Texas at Austin, 
1~University Station, C1400, Austin, TX, 78712}
\altaffiltext{3}{Dept.\ of Physics \& Astronomy, Johns Hopkins
University, 3400 N.\ Charles St., Baltimore, MD, 21218}

\slugcomment{Accepted for publication in ApJ Letters 2005--05--13}

\begin{abstract}

The disk around \aumic, an M1 star in the \bpic\ Moving Group, is 
extraordinarily well-suited for comparison to the \bp\ debris disk (type A5V).
We use far-UV absorption spectroscopy of \au\ to probe its edge-on
disk for small amounts of \hh, the primary constituent of gas giant planets.
Our conservative upper limit on the line-of-sight \hh\ column density is
$1.7 \times 10^{19}$~\cm, which is 18.5 times lower than the limit 
obtained from non-detection of sub-mm CO emission \citep{Liu:2004}.  
In addition, there is a hint of \hh\ absorption at a column density 
an order of magnitude or more below our upper limit.  
The \hh-to-dust ratio in the \au\ disk is $<$~6:1, 
similar to that in the \bp\ disk.
This shows that the primordial gas has largely been dissipated 
in less than about 12~Myr for both disks, despite their very different 
stellar masses and luminosities.
It is extremely difficult to form a giant planet around \au\ with 
current core-accretion scenarios in such a short time.

\end{abstract}



\section{Introduction} \label{sec:intro}

A circumstellar (CS) disk around a young star evolves from a massive remnant 
of star formation (a primordial disk, composed mostly of \hh\ gas) 
to a relatively low mass, gas-free planetary system.
But there are difficulties reconciling observations of CS disks with current theories of giant planet formation.
In the traditional picture, giant planets form by accretion of gas onto a 
massive solid core, which must form before the primordial gas in the disk dissipates. 
Core formation takes at least a few~Myr for solar-type stars, much longer for lower mass stars \citep{Laughlin:2004}.
Observations suggest that the primordial gas dissipation timescale is shorter 
than about 10~Myr, possibly as short as 1~Myr in regions of high-mass star formation \citep{Bally:1998}.
With these short gas lifetimes, it would be fairly difficult to form giant 
planets with current core-accretion scenarios around most solar-type stars 
and extremely difficult around all low-mass stars.

However, giant planets have been found around many G and K-type stars, 
and two M-type stars \citep[e.g.][]{Butler:2004}.
This has led some workers to suggest alternative theories for giant planet formation, i.e.\ that they form rapidly by direct collapse in 
gravitationally unstable disks \citep{Boss:1998}.
But it is not clear that current core-accretion models must be rejected, 
due to the uncertain timescale for primordial gas dissipation around stars of various masses in different star-forming environments.
Comparison between the abundance of giant planets around M-stars and the typical \hh\ gas lifetime in their disks will be a powerful constraint on theories of giant planet formation.

The M1 star \aumic\ (GJ~803) has long been studied as a nearby flare star 
($d = 9.94 \pm 0.13$~pc).
%
%
It is co-moving with the A5V star \bpic, indicating they formed in the same region and are the same age \citep[$12^{+8}_{-4}$~Myr;][]{Zuckerman:2001a}.
The CS dust around \au\ was imaged in reflected light and shown to lie in an 
edge-on disk \citep{Kalas:2004}.
This disk is uniquely suited for comparison to the \bp\ debris disk and 
provides a valuable opportunity to examine the effect of stellar mass 
on disk evolution.

The inclination of the \au\ disk within 50~AU of the star is less than 
$1^\circ$ \citep{Krist:2005}, even more edge-on than the \bp\ disk
\citep[$i = 3^\circ$;][]{Heap:2000}.
CS gas in the \bp\ disk has been observed with line-of-sight
absorption spectroscopy \citep[e.g.\ CO;][]{Roberge:2000}, and a
low upper limit on the \hh\ abundance was set using far-UV 
absorption spectroscopy \citep{LecavelierdesEtangs:2001}.
This limit showed that the primordial gas has largely been dissipated.
In this paper, we apply similar analysis to far-UV spectra of \au,
to probe sensitively for traces of primordial \hh\ gas.

\section{Observations} \label{sec:obs}

We used two sets of archival far-UV spectra of \au, one each from the 
\textit{Far Ultraviolet Spectroscopic Explorer} (\fuse) and the 
\textit{Hubble Space Telescope} Space Telescope Imaging Spectrograph (STIS).
Detailed descriptions of the data and analyses of the stellar 
emission lines may be found in \citet{Redfield:2002} and \citet{Pagano:2000}.
\au\ was observed with \fuse\ on 2000 August 26 and 2001 October 10 for 
17.3~ksec and 26.5~ksec, respectively.
These spectra cover \mbox{905 -- 1187~\AA} with a resolution
of about 15~\kms\ ($R = \lambda/\Delta \lambda = 20,000 \pm 2000$).
The data were recalibrated with CALFUSE version 2.4.0.
The STIS data were taken 1998 September 6 (exposure time = 10.1~ksec).
These spectra cover \mbox{1170 -- 1730~\AA} with a resolution of 4.4~\kms\ at 1350~\AA\ ($R = 68,000$).
We recalibrated the data with CALSTIS 2.17b.

Stellar flares occurred in both datasets.
Since emission line shapes and strengths can change during flares
\citep[e.g.][]{Redfield:2002}, data taken at these times were excluded 
from our analysis.
The \fuse\ bandpass is covered by 8 partially overlapping spectra,
two of which have the same zero-point wavelength offset (LiF1A and LiF1B).
The bandpasses of the LiF1B \fuse\ spectrum and the STIS spectrum overlap, 
so we set the absolute wavelength calibration of the LiF1A and LiF1B spectra 
by cross-correlating on the \ion{C}{3} $\lambda 1176$ stellar emission line 
seen in both datasets.
The zero-point offset of the \fuse\ spectra was small ($\approx 5$~\kms).

\section{Analysis} \label{sec:anal}

\begin{figure}[ht!]
\epsfig{file=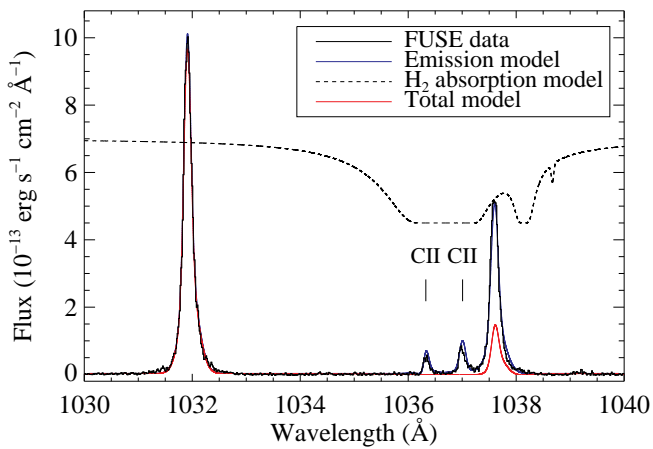,width=3.45in}
\caption{Portion of the \fuse\ LiF1A spectrum of \au\ with a 
demonstration of the effect of line-of-sight \hh\ absorption 
on the \Os\ $\lambda \lambda 1032, 1038$ and \Ct\ $\lambda \lambda 
1036, 1037$ emission doublets.
The intrinsic emission model, convolved with the \fuse\ LSF, is 
overplotted with a blue solid line.
A convolved \hh\ absorption model with the parameters of the upper limit 
from non-detection of sub-mm CO emission \citep[$N_\mathrm{total} = 3.15
\times 10^{20}$~\cm, $T = 40$~K;][]{Liu:2004} is plotted with a black 
dashed line.
For the \hh\ model shown, we assumed unresolved lines and set the velocity 
of the gas to the stellar velocity ($v_\star = -4.89$~\kms).
The total model is the absorption model multiplied by the emission model, 
then convolved with the \fuse\ LSF (red solid line).
The \Os\ $\lambda 1032$ emission line is little affected by \hh\ absorption,
but such a large \hh\ column density would dramatically suppress the 
\Os\ $\lambda 1038$ line and completely absorb both \Ct\ emission lines.
\label{fig:fuse} }
\end{figure}

Some of the many strong far-UV electronic transitions of \hh\ overlap with 
commonly seen stellar emission lines, i.e.\ the \ion{O}{6} 
$\lambda \lambda 1032, 1038$ doublet originating in the stellar transition 
zone, and the chromospheric \ion{C}{2} $\lambda \lambda 1036, 1037$ doublet.
The portion of the \fuse\ LiF1A spectrum including these lines is 
shown in Figure~\ref{fig:fuse}.
We have demonstrated that these emission lines can provide the background 
flux for absorption spectroscopy of \hh\ 
\citep{Roberge:2001, LecavelierdesEtangs:2001}.

\subsection{Creation of the Emission Line Model} \label{sub:emission}

The first step in analyzing the \fuse\ spectra for molecular hydrogen 
absorption was to model the intrinsic flux in the emission lines.
This is an emission model with infinite spectral resolution that will be multiplied by an intrinsic absorption model, then convolved with the 
instrumental line-spread function (LSF) to produce a model of the observed flux.

The \Os\ $\lambda 1032$ line is little affected by \hh\ absorption, and 
the doublet lines are always in the optically thin ratio of 2:1, 
unless the $\lambda 1038$ line is suppressed by \hh. 
The measured integrated flux ratio of the two \Os\ lines is $1.97 \pm 0.19$, 
immediately indicating that there is little line-of-sight \hh\ absorption. 
We began by least-squares fitting the sum of a narrow and 
a broad Gaussian to the $\lambda 1032$ line, since chromospheric and 
transition zone emission lines from an active star like \au\ are not well 
fit with single Gaussians \citep[e.g.][]{Redfield:2002}.
The widths and peak intensities of the Gaussians were adjusted to remove 
the effect of the Gaussian \fuse\ LSF.
The parameters from the fit to the $\lambda 1032$ line were used to model the $\lambda 1038$ line, with the peak flux set by the optically thin ratio.

\begin{figure}[ht!]
\epsfig{file=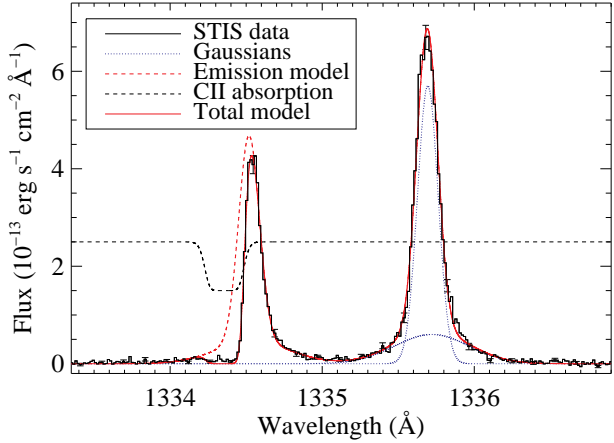,width=3.45in}
\caption{Portion of the E140M STIS spectrum of \au\ showing the
\Ct\ $\lambda 1335$ emission triplet (analyzed to aid in modeling
of the \fuse\ \Ct\ emission lines).
The blue wing of the \Ct\ $\lambda 1334.5$ line is cut off by line-of-sight 
\Ct\ absorption.
The narrow and broad Gaussians from our modeling of the \Ct\ $\lambda 
1335.7$ line are overplotted with dotted blue lines.
The emission line model of the triplet, convolved with the STIS LSF, is overplotted with a red dashed line.
The convolved \Ct\ $\lambda 1334.5$ absorption model is plotted with a 
black dashed line.
The total model is the absorption model multiplied by the emission model, 
then convolved with the STIS LSF (solid red line).
\label{fig:stis} }
\end{figure}

The \Ct\ $\lambda \lambda 1036, 1037$ lines provide a tighter constraint 
on \hh\ absorption, but were difficult to model, since both can be affected.
Therefore, we analyzed the \Ct\ $\lambda 1335$ triplet in the STIS data, 
shown in Figure~\ref{fig:stis}. 
Examination of the line profiles shows that the blue wing of the 
$\lambda 1334.5$ emission line is cut off by ground-state \Ct\ absorption, 
which was initially identified as interstellar (IS), but the $\lambda 1335.7$ emission line is not affected by excited-state \Ct\ absorption.

We found the parameters of the \Ct\ emission by analysis of the 
$\lambda 1335.7$ line, which is blended with the weak $\lambda 1335.6$ line.
The emission was modeled as the sum of a narrow and broad Gaussian
for each line, setting the line separation from the rest wavelengths.
The relative line strengths were set using the products of the upper state multiplicities and Einstein A-coefficients.
We then performed $\chi^2$ minimization.
The widths and peak intensities from this analysis were adjusted to
remove the effect of the STIS LSF, which is very small since the lines are 
well-resolved. 
The narrow component is at $v_\mathrm{n} = -3.0 \pm 3.4$~\kms, where 
the error bar is the $\pm 1 \sigma$ accuracy of the STIS wavelength 
calibration after a standard target acquisition \citep{Pagano:2000}.
Within the uncertainty, the \Ct\ emission is at the velocity of the star
\citep[$v_\star = -4.89 \pm 0.02$~\kms;][]{BarradoyNavascues:1999}.
As is usually the case, the broad component is slightly redshifted
($v_\mathrm{b} = 3.0 \pm 3.4$~\kms).
The narrow component has $FWHM = 35.63 \pm 0.55$~\kms\ and the
broad component has $FWHM = 129.4 \pm 3.9$~\kms.

The parameters from the $\lambda 1335.7$ minimization were used 
to model the $\lambda 1334.5$ emission line, again setting the wavelength 
separation and relative line strength from the atomic data.
We found that the $\lambda 1335.7$ line is not optically thin,
since the peak intensities of the Gaussians in the $\lambda 1334.5$ line 
had to be increased slightly to fit the blue wing of the line.
The emission model of the triplet convolved with the STIS LSF is shown with a dashed red line in Figure~\ref{fig:stis}.

To find the parameters of the \Ct\ $\lambda 1334.5$ absorption, we used a 
Voigt intrinsic line profile to create an absorption model.
A total model was made by multiplying the triplet emission model by
the absorption model, then convolving the result with the STIS LSF.
The absorption parameters from $\chi^2$ minimization of the total model are $N = 7.9^{+1.9}_{-1.0} \times 10^{14}$~\cm\ (column density), 
$b = 17.50 \pm 0.63$~\kms\ (Doppler broadening parameter), and 
$v = -39.5 \pm 3.4$~\kms.
The convolved final model for the \Ct\ $\lambda 1335$ triplet is overplotted
with a solid red line in Figure~\ref{fig:stis}.
The \Ct\ absorption is broad and blueshifted with respect to 
both the star and other absorption lines seen toward \au\ 
\citep[e.g.\ \ion{D}{1} at $v = -21.7$~\kms;][]{Redfield:2004}.
This suggests that the line-of-sight \ion{C}{2} absorption may not 
be primarily IS.

Next, we modeled the \Ct\ emission doublet in the \fuse\ spectrum, 
using the velocities and widths of the narrow and broad Gaussians 
and the ratio of their peak intensities from the STIS analysis.
The separation of the two lines and their relative strengths 
were calculated from the atomic data.
The doublet emission model was multiplied by a \Ct\ $\lambda 1036$ 
absorption line, using the parameters given in the previous paragraph.
The remaining free parameter was a scaling factor on the peak intensity 
of the $\lambda 1036$ line, which was determined by matching the model to 
the blue wing of the line.
We found that the ratio of the line strengths calculated from the atomic 
data had to be adjusted slightly to fit the blue wing of the $\lambda 1037$ 
line, again indicating it is not optically thin.

The \Ct\ and \Os\ models were added together to produce an infinite resolution emission model for the 1030 -- 1040~\AA\ region of the \fuse\ spectrum.
This model convolved with the \fuse\ LSF is plotted in Figure~\ref{fig:fuse} 
with a solid blue line.
One can immediately see that the \Ct\ emission lines in the \fuse\ data
are slightly narrower than expected from the width of the STIS \Ct\ lines.
There is no intrinsic reason for this, since we excluded flares from 
both data sets.
Self-absorption suppresses the peak flux of an emission line, but does not 
make it appear narrower.
The emission lines are resolved, so the uncertainty in the spectral 
resolution of the \fuse\ data has little effect on the model.
Other obvious \fuse\ instrumental or data artifacts tend to make 
emission lines appear broader than they really are, not narrower.
It seems there are only two ways to make the \fuse\ \Ct\ lines appear
narrower than the STIS lines.
The first is a calibration error that artificially broadened the STIS lines.
There is such an error in STIS far-UV time-tag data (like ours) 
retrieved from the archive before 2004 April 21\footnote{see
\url{http://www.stsci.edu/hst/stis/documents/newsletters/stan0412.html}}. 
However, we recalibrated our data with the corrected pipeline software.
The other way is to remove some emission from the \fuse\ lines with 
superimposed \hh\ absorption (discussed further in \S\ref{sub:detect}).

\subsection{Molecular Hydrogen Model Fitting} \label{sub:h2}

An \hh\ absorption model was created using Voigt line profiles 
and molecular data from \citet{Abgrall:1993a}.
The free parameters were $N(J=0)$, the column density in the
$\nu = 0, \ J = 0$ level, $N(J=1)$, the column density in the
$\nu = 0, \ J = 1$ level, and $v$, the velocity of the gas. 
Only the two lowest energy levels were included, since they contain more than 
$90\%$ of the molecules at the typical temperatures of debris disks 
($< 200$~K). 
The intrinsic width of any absorption is unknown, so we made the
most conservative assumption (unresolved lines) and set the
Doppler broadening parameter to be very small ($b = 1$~\kms).
The \hh\ model was multiplied by the \fuse\ emission model, then convolved 
with the \fuse\ LSF.
The resulting total model was compared to the data between 1035.5~\AA\ and 
1037.7~\AA\ (the region most affected by \hh\ absorption).
To ensure that all local minima were found, $\chi^2$ 
minimization was performed using very large ranges of parameters
($10^{12} \ \mathrm{cm}^{-2} \ \leq \ N(J=0) \ \mathrm{and} \ N(J=1)
\ \leq 10^{22} \ \mathrm{cm}^{-2}, \ -60 \ \mathrm{km \ s}^{-1} \ \leq \
v \ \leq \ + 60 \ \mathrm{km \ s}^{-1}$).

\section{Results}

\subsection{Upper Limits on \hh} \label{sub:upper}

\begin{figure}[ht!]
\epsfig{file=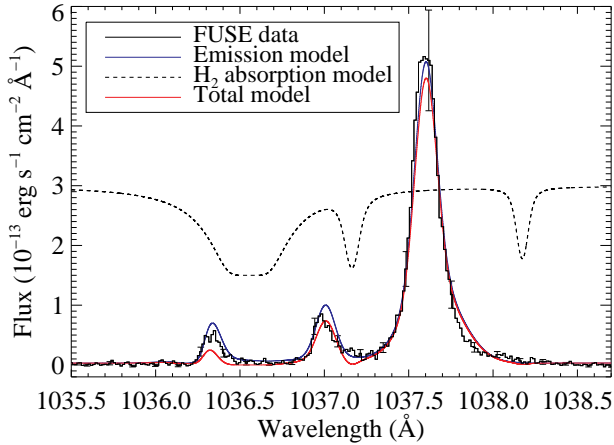,width=3.45in}
\caption{Upper limit on line-of-sight \hh\ absorption toward \au.
The portion of the \fuse\ LiF1A spectrum near the \ion{C}{2} 
$\lambda \lambda 1036, 1037$ doublet is shown, with the convolved 
emission model overplotted (blue solid line).
The black dashed line shows the convolved upper limit \hh\ absorption model
($N(J = 0) = 1.6 \times 10^{19}$~\cm, $N(J = 1) = 6.3 \times 10^{17}$~\cm).
The total model is the absorption model multiplied by the emission model,
then convolved with the \fuse\ LSF (solid red line).
For the \hh\ model shown, the velocity was set to $+ 4.0$~\kms. 
This is the most redshifted velocity expected for line-of-sight 
CS gas in Keplerian rotation in the edge-on disk, given the radial 
velocity of the star ($v_\star = -4.98$~\kms) and the combined accuracy 
of the \fuse\ and STIS wavelength calibrations ($\pm 8.7$~\kms).
\hh\ absorption models with velocities blueward of $+ 4.0$~\kms\ have 
an even more dramatic effect on the \Ct\ emission lines.
\label{fig:h2} }
\end{figure}

Using our contours of $\chi^2$, we set upper limits on the line-of-sight \hh\ 
column densities of $N(J = 0) < 1.6 \times 10^{19} \ \mathrm{cm}^{-2}$ and
$N(J = 1) < 6.3 \times 10^{17} \ \mathrm{cm}^{-2}$.
These column densities are ruled out at the $\geq 6 \sigma$ level for all $v$.
In Figure~\ref{fig:h2}, the convolved \hh\ absorption model with these values is shown, and the total model overplotted on the data near the \Ct\ lines.
At lower column densities, the contours of $\chi^2$ are complicated 
by the presence of a local minimum in addition to the absolute minimum,
which is why we set the limits more conservatively from the $6 \sigma$ 
contour rather than the typical $3 \sigma$ contour.
The upper limit on the total line-of-sight \hh\ column density toward \au\ is 
$N_\mathrm{total} \, < \, 1.7 \times 10^{19} \ \mathrm{cm}^{-2}$;
we did not assume a gas temperature to get this limit. 
If one assumes that the \hh\ energy levels are thermally populated and 
fixes the gas temperature at any value below a few thousand K, the total
upper limit decreases.

From sub-mm observations, \citet{Liu:2004} placed a $3 \sigma$ upper limit 
on the CO column density in the \au\ disk of $6.3 \times 10^{13}$~\cm, 
assuming any gas was in thermal equilibrium with the observed dust 
($T = 40$~K).
Using a \cohh\ ratio of $10^{-7}$, this corresponded to a total \hh\ 
column density of $6.3 \times 10^{20}$~\cm.
The \cohh\ value is unusually low compared to the typical IS value, since the authors tried to take into account the fact that the \au\ disk is optically thin to IS UV photons.
To convert this column density from emission observations to a 
line-of-sight absorption column, we divided the sub-mm upper limit by 2.
This gives the lowest possible value, by assuming that the radial density distribution of the gas is extremely centrally peaked. 
Our $\chi^2$ contours show that the sub-mm upper limit on the line-of-sight
\hh\ column density is ruled out at the $> 20 \sigma$ level for all $v$.
Such a large column density would dramatically suppress the \Os\ 
$\lambda 1038$ emission line and completely obliterate both \Ct\ 
emission lines (see Figure~\ref{fig:fuse}).
Our upper limit in the previous paragraph is 18.5 times lower than the sub-mm upper limit, does not assume a gas temperature, and is free from the large uncertainties introduced by assuming a \cohh\ ratio.

\subsection{Possible Line-of-Sight \hh\ Absorption} \label{sub:detect}

The fact that the \fuse\ \ion{C}{2} emission lines are narrower than 
the STIS lines suggests there is a small amount of \hh\ absorption toward \au.
The minimization with fixed $b$ described above showed a second minimum 
in the $\chi^2$ contours at \hh\ column densities about an order of magnitude 
below our $6 \sigma$ upper limits.
In addition, there have been claims of weak fluorescent \hh\ emission 
lines in the \fuse\ and STIS spectra \citep{Pagano:2000,Redfield:2002};
however, these lines are only detected at the $\approx 2 \sigma$ level.

We attempted to detect line-of-sight \hh\ absorption by performing $\chi^2$ minimization with all parameters free, including $b$.
These contours showed a minimum at large $b$ and column densities a few orders 
of magnitude below the upper limits in \S\ref{sub:upper}.
But as before, there were multiple minima in the $\chi^2$ contours, indicating there are multiple sets of parameters producing \hh\ models which 
fit the data about equally well.
Therefore, the model is overdetermined given the quality of the data.
But taken at face value, our analysis indicates that the putative line-of-sight \hh\ absorption is broad and blueshifted with respect to the star, as was the line-of-sight \ion{C}{2} absorption in the STIS data.

\section{Discussion} \label{sec:disc}

\citet{Liu:2004} measured a sub-mm dust mass of $0.011 \ M_\earth$
within 70~AU of \au.
Their upper limit on the \hh\ mass ($1.3 \ M_\earth$) gives
a limit on the \hh-to-dust ratio of $<$~118:1.
Our limit on the line-of-sight \hh\ column density reduces the possible gas 
mass by at least a factor of 18.5, giving an upper limit on the \hh-to-dust 
ratio of $<$~6:1.
The typical IS ratio is 100:1;
therefore, the \hh\ gas in the \au\ disk has been depleted relative to the dust.
The primordial gas is largely dissipated, and the gas lifetime was less 
than about 12~Myr.

\citet{Laughlin:2004} showed that a core massive enough to gravitationally 
accrete a large gaseous envelope could not form within 10~Myr around a $0.4~M_{\sun}$ star.
The only way current core-accretion scenarios can form a giant planet around 
an M-star is if its primordial gas survives much longer than about 10~Myr.
Therefore, it is unlikely that a giant planet could form around \au\ 
through these scenarios. 
There is indirect evidence of a giant planet around \au, since the 
central region of the disk is cleared of dust grains \citep{Liu:2004}
and the disk appears slightly warped in the \hst-ACS scattered light 
image \citep{Krist:2005}.
Central holes and warps have been attributed to the dynamical influence 
of an unseen giant planet.

We now consider the case of \bp.
The disk dust mass is $0.04 \ M_\earth$ \citep{Dent:2000}.
Using the upper limit on the \hh\ gas mass from \fuse\ 
\citep[$\lesssim 0.1 \ M_\earth$;][]{LecavelierdesEtangs:2001}, the \bp\ 
\hh-to-dust ratio is $<$~3:1, similar to the limit for \au.
This indicates that the primordial gas lifetimes in the two systems were 
both quite short, despite the very different stellar masses and luminosities 
($L_\mathrm{AU~Mic} = 0.1 \ L_\sun; \ L_\mathrm{\beta Pic} = 8.7 \ L_\sun$).
There is no O or B star in the \bp\ Moving Group, so the short lifetimes 
are not easily explained by external irradiation.
Given their spectral types, photoevaporation by the central stars seems 
unlikely to produce short gas lifetimes in both systems.
Other possible dissipation mechanisms include stellar winds, 
magnetospheric accretion, and planet formation.

Our observation of broad, blueshifted line-of-sight \Ct\ absorption
(and possibly \hh) would lead one to suspect that there is outflowing, 
low-ionization gas near the star, possibly in a stellar or disk wind.
However, there may be an unusually large number of blended IS 
components at different redshifts along the sight line to \au.
We plan to investigate further by re-examining the putative fluorescent
\hh\ emission lines and all atomic absorption features in the STIS spectrum 
previously identified as IS.



\end{document}